\documentclass[aps,prd,twocolumn,nofootinbib,floatfix]{revtex4-2}

\usepackage[T1]{fontenc}
\usepackage{lmodern}
\usepackage{amsmath,amssymb,amsfonts,mathtools,bm}
\usepackage{graphicx}
\usepackage{hyperref}
\usepackage{xcolor}
\usepackage{physics}
\usepackage{booktabs}
\usepackage{microtype}

\definecolor{linkblack}{RGB}{20,20,20}
\definecolor{citeblue}{RGB}{0,70,140}
\definecolor{urlblue}{RGB}{0,90,120}
\usepackage{float}
\hypersetup{
  colorlinks=true,
  linkcolor=linkblack,
  citecolor=citeblue,
  urlcolor=urlblue
}

\newcommand{\mpl}{M_{\rm Pl}}
\newcommand{\rc}{r_c}
\newcommand{\Vpot}{V}
\newcommand{\Ucal}{\mathcal U}
\newcommand{\Hc}{\mathcal H}

\newcommand{\safeincludegraphics}[2][]{%
  \IfFileExists{#2}{\includegraphics[#1]{#2}}{\fbox{\parbox[c][0.25\textheight][c]{0.9\linewidth}{\centering Figure placeholder:\\ \texttt{\detokenize{#2}}}}}%
}

\begin{document}

\title{Geodesically Complete Curvature-Bounce Inflation}

\author{Damien A. Easson}
\email{easson@asu.edu}
\affiliation{Department of Physics, Arizona State University, Tempe, Arizona 85287, USA}
\affiliation{Beyond Center for Fundamental Concepts in Science, Arizona State University, Tempe, Arizona 85287, USA}

\begin{abstract}
The early universe need not be described by an incomplete inflationary phase
connected to a separate, more exotic prehistory. Recent results show that,
within non-static FRW cosmology, only positive spatial curvature permits a
nonsingular, geodesically complete universe with ANEC-respecting matter. We
construct a geodesically complete closed $k=+1$
bounce-plus-inflation cosmology in ordinary general relativity, sourced by a
single canonical scalar field with a positive vacuum offset. The bounce is
supported by curvature rather than exotic stress energy: the matter content
satisfies the NEC throughout and violates only the strong energy condition, as
in any accelerated expansion. The solved branch remains sub-Planckian and evolves
onto a curvature-diluted slow-roll phase with inflationary observables
consistent with current constraints. The pivot-scale predictions are
$n_s=0.9617$, $r=0.0045$ at $N_*=55$ and $n_s=0.9650$, $r=0.0037$ at
$N_*=60$. Direct evolution of closed-universe infrared perturbations shows
regular tensor and scalar propagation through the bounce and inflationary era,
with the physical curvature perturbation freezing in the standard way. This
gives a minimal explicit realization of a complete early-universe cosmology in
the closed FRW branch selected by completeness and ANEC compatibility.
\end{abstract}

\maketitle

\section{Introduction}
\label{sec:intro}

Inflation explains the observed near-scale-invariant spectrum of primordial
perturbations and the large-scale homogeneity of the universe
\cite{Guth:1980zm,Linde:1981mu,Albrecht:1982wi,Planck:2018vyg}, but by itself it
does not settle the status of the earliest cosmic history. In the standard
picture, inflationary spacetimes are past-incomplete, so a satisfactory account
of the universe still appears to require either a beginning or a pre-inflationary
phase governed by additional physics \cite{Borde:2001nh}. This expectation has encouraged a large literature on singularity resolution through modified
gravity, higher derivatives, nonminimal couplings, multifield dynamics,
Horndeski-type sectors, loop-quantum-cosmology effects, string-motivated
constructions, and related extensions of the standard low-energy framework (see, e.g.
\cite{Khoury:2001wf,Novello:2008ra,Easson:2011zy,Cai:2012va,Battefeld:2014uga,Brandenberger:2016vhg,Creminelli:2016zwa,
Ijjas:2016tpn,Ijjas:2016vtq,Dobre:2017pnt,Cai:2016thi,Cai:2017tku,Kinney:2023urn}).

We argue that expectation is too broad. The question is not whether one can
engineer singularity resolution, but which Friedmann--Robertson--Walker (FRW) models can support a nonsingular,
geodesically complete, non-static cosmology while remaining compatible with a consistent quantum field theory description of matter. Recent work has explored the issue considerably \cite{Lesnefsky:2022fen,Geshnizjani:2023hyd,Easson:2024fzn,Easson:2024uxe,GarciaSaenz:2024gcbi}. Within general relativity, geodesically complete FRW cosmologies with nontrivial time evolution require a period of accelerated expansion, and the explicit complete examples known to date exhibit bouncing behavior \cite{Easson:2024uxe,Easson:2024fzn}. In addition, geodesic completeness
together with averaged null energy condition (ANEC) compatibility singles out the closed $k=+1$ family: flat and
open non-static FRW universes cannot be simultaneously nonsingular,
geodesically complete, and ANEC-consistent, whereas closed universes can \cite{Burwig:2025hrr, Burwig:2026fsy}.  Positive
spatial curvature is therefore singled out by completeness and semiclassical
energy conditions.

The ANEC requires that
$\int T_{\mu\nu} k^\mu k^\nu\, d\lambda \ge 0$ along complete null geodesics,
where $T_{\mu\nu}$ is the stress-energy tensor, $k^\mu$ is a null tangent
vector, and $\lambda$ is an affine parameter. Unlike the pointwise energy
conditions, ANEC is compatible with known quantum effects and is widely
regarded as a robust semiclassical constraint
\cite{Wald:1991xn,PhysRevD.51.4277,Graham:2007va,Hartman:2016lgu,Wall:2009wi}.

The above observations change the status of the model-building problem. Closed
inflationary universes and models with \(\Omega>1\) have been studied
previously, including quantum-creation and finite-inflation scenarios
\cite{Linde:1995xm,Linde:2003hc}. Earlier work has also shown that positive
spatial curvature can support nonsingular bounce-plus-inflation cosmologies in
Einstein gravity with scalar matter, and that such models can be
observationally viable
\cite{Ellis:2002we,Ellis:2003qz,Graham:2011nb,Matsui:2019}. Related
curvature bounces catalyzed by vacuum energy or dark energy have  been
analyzed in anisotropic settings \cite{Bramberger:2019darkbounce}, as well.

Our goal is to ask whether the closed FRW sector selected by geodesic completeness and
ANEC compatibility admits an explicit bounce realization within ordinary general
relativity, and whether that realization remains perturbatively well behaved on
its exact background solution. In short, can one build a complete model of the early universe that satisfies observational constraints within the well-known and empirically verified laws of physics? Earlier analyses of perturbations through closed or general-relativistic
bounces include, for example,
Refs.~\cite{Hwang:2001nonsingular,Gordon:2002grbounce,Martin:2003sf,Peter:2004bouncing}.
The novelty here is not perturbation propagation through a closed bounce by
itself, but the simultaneous realization of geodesic completeness,
ANEC-compatible canonical matter, viable slow-roll observables, and direct
infrared scalar-and-tensor regularity on one and the same closed-FRW
background.

In this work the bounce is
curvature-supported, not phantom-supported. No null-energy-condition violation
is required. No element from modified gravity is introduced. No higher derivatives,
nonminimal couplings, multifield dynamics, or quantum-gravity ingredients are
needed. The matter sector satisfies the NEC and ANEC throughout and violates
only the strong energy condition, exactly as happens in any phase of accelerated
expansion. The model therefore realizes, within minimal physics, the criteria that
the completeness analysis had already isolated on general grounds.

This simplicity here is more than merely aesthetic. Our natural target is the least
elaborate framework capable of providing a smooth past-eternal origin for
inflation, accounting for the observed early universe, and remaining within
well-understood dynamics. These 
demands do not conflict: they can be met simultaneously in a closed FRW
universe with ordinary canonical matter.
In such a history there is no first moment at which the effective description
must be initialized by a singularity, an instanton, or a separate
pre-inflationary boundary condition; the cosmology is instead a single smooth
solution defined for all cosmic time.

The model we study is everywhere regular, with bounded curvature invariants and
sub-Planckian evolution on the solved branch. Its observable predictions are
not generated at a bounce itself but on a later curvature-diluted slow-roll
branch, where the total expansion from the bounce to the end of inflation is
about $100$ e-folds and the standard pivot window lies comfortably within the
solved expanding history. Because observations probe only a narrow interval of the inflationary plateau,
the significance of the construction does not rely on the uniqueness of a finely tuned
global potential. Indeed, many nearby potentials can reproduce the same measured
$(n_s,r)$. The point is that a minimal, geodesically complete closed-FRW
background can realize such a plateau within a fully regular cosmological
history.

We test the perturbative consistency of the construction in the place most
sensitive to the bounce. Direct evolution of the closed-universe infrared tensor
and scalar modes shows regular propagation through the bounce and inflationary
era, with no pathology in the scalar canonicalization factor. The physical
curvature perturbation $\zeta_n$ freezes during inflation exactly as expected.
These directly evolved infrared modes should therefore be interpreted as
regularity diagnostics of the explicit curvature-bounce background, while the
observable CMB pivot predictions are controlled by the later slow-roll phase.
Any direct imprint of the bounce, if present at all, is expected only in the
extreme infrared, where cosmic variance dominates and severely limits discriminating power.

The remainder of this paper is organized as follows. In Sec.~\ref{sec:curvature_selection} we
review the setup and explain why positive curvature, inflation, and a
positive vacuum offset are the natural ingredients singled out by completeness
arguments. In Sec.~\ref{sec:model} we present the explicit curvature-bounce
inflationary construction, specify the implemented production branch, and
summarize its global evolution, regularity, and EFT control. In
Sec.~\ref{sec:observables} we derive the inflationary observables on the
slow-roll branch. In Sec.~\ref{sec:perturbations} we analyze linear
perturbations and show regular infrared tensor and scalar evolution through the
bounce and inflationary era, together with the proper inflationary freeze of the
physical curvature perturbation. The appendices record the exact generator used
for the production branch, the closed-universe scalar perturbation system and
numerical implementation, numerical robustness checks, a compact homogeneous
shear estimate, and a direct asymptotic completeness check for the explicit
background. We conclude in Secs.~\ref{sec:discussion}
and~\ref{sec:conclusion}.

\section{Curvature selection and bounce conditions}
\label{sec:curvature_selection}

A geodesically complete, smooth non-trivial FRW cosmology is highly
constrained. Although flat FRW cosmologies are observationally favored, they
cannot be simultaneously nonsingular and ANEC-compatible in the non-static
geodesically complete setting, and avoiding an initial singularity generally
requires violation of the classical energy conditions
\cite{Penrose:1964wq,Hawking:1966sx,Hawking:1966jv,Hawking:1967ju,Hawking:1970zqf,Hawking:1973uf,Burwig:2025hrr, Burwig:2026fsy}.

By contrast, closed \(k=+1\) universes admit curvature-supported bounces with
ordinary canonical matter, satisfying the NEC and ANEC while violating only
the SEC \cite{Ellis:2002we,Ellis:2003qz,Graham:2011nb,Matsui:2019,Easson:2024uxe,Burwig:2025hrr,Burwig:2026fsy}. This point is clearly visible in the closed FRW background equations. For
\begin{equation}
 ds^2=-dt^2+a(t)^2\,d\Omega_3^2,
\label{eq:frw_metric}
\end{equation}
with a canonically normalized scalar field \(\phi(t)\) and scalar potential
\(V(\phi)\), the homogeneous equations are
\begin{align}
 H^2 &= \frac{1}{3}\left(\frac{1}{2}\dot\phi^2+V(\phi)\right)-\frac{1}{a^2},
\label{eq:friedmann_closed}\\
 \dot H &= -\frac{1}{2}\dot\phi^2+\frac{1}{a^2},
\label{eq:hdot_closed}\\
 \ddot\phi &+3H\dot\phi+V_{,\phi}=0.
\label{eq:kg_closed}
\end{align}
Here and throughout, \(V(\phi)\) denotes the full scalar potential appearing in
the solved background, including the positive vacuum offset; equivalently one
may write \(V(\phi)=U(\phi)+\Lambda\) with \(\Lambda>0\). In our implemented
production branch, the stable minimum satisfies
\(V(\phi_{\min})=\Lambda\), so the late-time de Sitter vacuum energy is encoded
directly by this positive offset.

At a bounce, \(H=0\) and \(\dot a=0\), while \(\ddot a>0\). Using
\(\ddot a/a=H^2+\dot H\), the bounce condition becomes
\begin{equation}
 \left.\frac{\ddot a}{a}\right|_{\rm b}
 =
 -\frac{1}{6}\left(\rho_{\rm tot}+3p_{\rm tot}\right)_{\rm b}
 =
 -\frac{1}{3}\dot\phi_{\rm b}^2+\frac{1}{3}V_{\rm b}
 >0,
\label{eq:bounce_condition}
\end{equation}
where
\begin{equation}
 \rho_{\rm tot}=\frac12\dot\phi^2+V,
 \qquad
 p_{\rm tot}=\frac12\dot\phi^2-V.
\label{eq:rhop_total}
\end{equation}
Thus a closed-universe bounce requires violation of the strong energy
condition, \(\rho_{\rm tot}+3p_{\rm tot}<0\), but it does \emph{not} require
NEC violation. Indeed,
\begin{equation}
 \rho_{\rm tot}+p_{\rm tot}=\dot\phi^2\ge 0,
\label{eq:nec_scalar}
\end{equation}
so the NEC is automatically satisfied, while the curvature term in
Eq.~\eqref{eq:hdot_closed} supplies the positive contribution needed to make
\(\dot H_{\rm b}>0\) (a tell-tale signal of NEC violation in the flat $k=0$ FRW model).

This is the essential mechanism behind the construction: positive spatial
curvature, rather than exotic stress energy, supports the bounce. The matter
content is a canonical scalar field in ordinary general relativity, so
the NEC is preserved throughout; the only pointwise energy condition that must
be violated is the SEC, as in any inflationary model.

The same closed geometry also naturally accommodates a stable positive vacuum
offset. In the empty closed-FRW limit, this is simply global de Sitter space. For a closed FRW vacuum one has
\(
H^2=\Lambda/3-a^{-2},
\)
so a positive vacuum offset is the natural way to maintain a nontrivial
late-time vacuum branch with \(H^2\ge 0\). In this geometric sense, a
nontrivial closed-FRW vacuum branch requires a positive vacuum contribution.
The present model may therefore be viewed as a minimal nontrivial realization
of that lesson: a curvature-supported nonsingular bounce followed by ordinary
slow-roll inflation, graceful exit, and a stable minimum.

\section{Construction of the curvature-bounce inflationary model}
\label{sec:model}

\subsection{Construction strategy}

Our model is built by specifying a smooth
closed-FRW background history and then reconstructing the canonical scalar
dynamics from the background equations given in
Sec.~\ref{sec:curvature_selection}. As opposed to starting from a
closed-form potential and then solving forward, the background history is taken
as primary, and the corresponding scalar profile and potential are obtained from
Eqs.~\eqref{eq:friedmann_closed}--\eqref{eq:kg_closed} on the solved branch.

\begin{figure}[t]
\safeincludegraphics[width=\columnwidth]{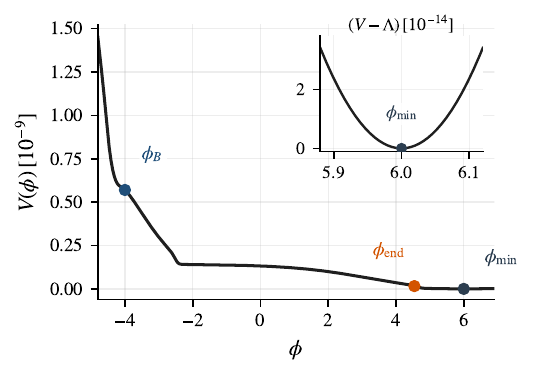}
\caption{
Scalar potential \(V(\phi)\) for the curvature-bounce inflationary model.
The marked field values indicate the bounce region, the end of inflation, and
the stable minimum. The positive offset allows the same smooth canonical-scalar
potential to support the nonsingular bounce, the inflationary plateau, and the
late-time vacuum structure. The observable CMB window lies on the flat
inflationary plateau. The
inset shows \(V(\phi)-\Lambda\) near the stable minimum on a linear scale,
making the local quadratic structure visible; the full potential has
\(V(\phi_{\min})=\Lambda\).
}
\label{fig:potential}
\end{figure}

The potential used for the production branch has a clear construction. Near the
bounce, it is anchored to the exact canonical-scalar reconstruction of the
simple closed quadratic curvature-bounce family,
\(a(t)\sim t^2+c\), for a constant $c>0$. This prototype belongs to the class of complete
accelerating models with \(\ddot a>0\) for all \(t\) discussed in
\cite{Easson:2024fzn}. In the production implementation, the same bounce family
is rewritten in terms of the bounce scale \(a_b=1/(2A)\) and provides the
bounce-side potential responsible for the curvature-supported bounce. This is
then smoothly joined to a Starobinsky-like inflationary plateau
\cite{Starobinsky:1980te}, followed by a smooth transition to a reheating and
stabilized minimum chosen to yield an observationally viable post-bounce
slow-roll branch; see Fig.~\ref{fig:potential}. The exact interpolation
functions and component potentials are given in
App.~\ref{app:production_generator}. The solved background is therefore a
controlled composite: an exact curvature-bounce potential near the bounce,
written in production-branch variables, smoothly glued to a post-bounce
slow-roll inflationary branch chosen to match observations.

In practice, the presented model is generated
numerically and stored as the background arrays
\begin{equation}\label{eq:arrays}
    \{t,\eta,a,H,\phi,\dot\phi\}.
\end{equation}
These arrays are then treated as the input for the observables and
perturbation analyses. This is why the appendix reconstructs \(V\),
\(V_{,\phi}\), and \(V_{,\phi\phi}\) directly from the solved background rather
than from an independent closed-form potential.

\subsection{Implemented production branch}

We use the amplitude-normalized fiducial solution, obtained
from the original benchmark family by the rescaling
\begin{equation}
V_0\to \lambda V_0,\qquad
m^2\to \lambda m^2,\qquad
\Lambda\to \lambda\Lambda,\qquad
A\to \sqrt{\lambda}\,A,
\label{eq:family_rescaling}
\end{equation}
with \(\lambda\simeq 0.475\). This rescaling serves to preserve the qualitative
bounce-plus-inflation structure and the slow-roll shape observables while
bringing the scalar amplitude into the observed range. The exact analytic generator defining the benchmark solution, including the
smooth interpolation functions, the full potential Eq.~\eqref{eq:Vtotal_prod_appendix}, the
bounce initial data, and the numerical integration procedure, is given in
App.~\ref{app:production_generator}, while Table~\ref{tab:model_parameters} lists the
corresponding implemented parameter values.

The value of \(\lambda\) was fixed by matching the first-order
slow-roll estimate \(A_s\simeq H_*^2/(8\pi^2\epsilon_1)\) at the benchmark point
\(N_*=55\) to the observed CMB scalar amplitude,
\(A_s\simeq 2.1\times10^{-9}\) \cite{Planck:2018vyg}.

\begin{table}[t]
\caption{Implemented parameters for the amplitude-normalized production branch.}
\label{tab:model_parameters}
\begin{ruledtabular}
\begin{tabular}{lc}
Parameter & Value \\
\hline
$\lambda$ & $0.47497$ \\
$A$ & $6.892\times 10^{-6}$ \\
$\phi_0$ & $-4.0$ \\
$\Delta_L$ & $0.8$ \\
$\gamma$ & $1.5$ \\
$\phi_{\rm trans}$ & $-2.5$ \\
$\delta_1$ & $0.10$ \\
$V_0$ & $1.425\times 10^{-10}$ \\
$\beta$ & $\sqrt{2/3}$ \\
$\phi_p$ & $4.0$ \\
$\phi_{\rm reh}$ & $4.6$ \\
$\delta_2$ & $0.15$ \\
$\phi_m$ & $6.0$ \\
$m^2$ & $4.750\times 10^{-12}$ \\
$\Lambda$ & $4.750\times 10^{-121}$ \\
\end{tabular}
\end{ruledtabular}
\end{table}

\begin{figure}[t]
\safeincludegraphics[width=\columnwidth]{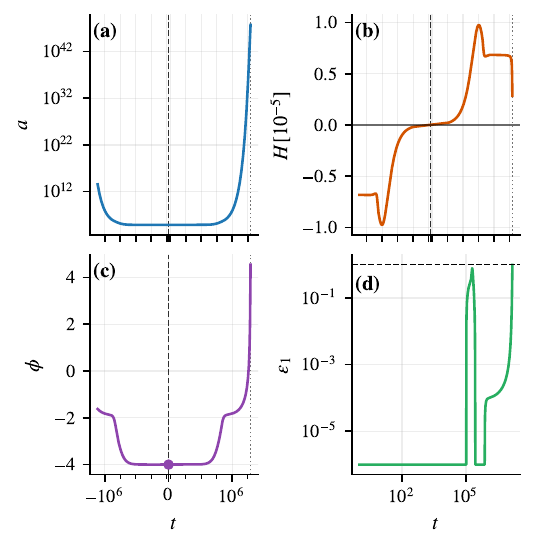}
\caption{
Background evolution of the curvature-bounce inflationary model over a finite
portion of the solved history used in the numerical analysis. Panels (a)--(d)
show, respectively, the scale factor \(a(t)\), the Hubble parameter \(H(t)\),
the scalar-field history \(\phi(t)\), and the first Hubble slow-roll parameter
\(\epsilon_1\) on the expanding branch. The closed \(k=+1\) universe undergoes
a smooth nonsingular bounce, evolves through a long period of accelerated
expansion, and later exits inflation. Panel (c) makes clear that the bounce
and subsequent inflationary phase are realized by a single smooth
canonical-scalar history on the explicit solved branch. The full asymptotic de
Sitter behavior described in the text lies beyond the plotted interval.
}
\label{fig:background}
\end{figure}

\subsection{Global evolution of the scalar field}

An important conceptual point is that the scalar field does not ``start'' at any finite initial time. In a geodesically complete evolution, there is no such first moment. Rather,
\(\phi(t)\) is a smooth global field configuration defined on the entire
spacetime, \(t\in(-\infty,\infty)\). In the explicit solved background,
\(\dot\phi\) vanishes at the bounce, so \(\phi(t)\) has a local extremum
there. The left-right orientation of Fig.~\ref{fig:potential} is therefore
conventional: what matters physically is the existence of a single smooth global scalar
history compatible with the solved background, not monotonicity through the
bounce.

Fig.~\ref{fig:background}$(c)$ shows only a finite portion
of the full evolution of \(\phi(t)\); the eternal asymptotic behavior lies outside the
plotted interval. In particular, the remote
contracting de Sitter regime, in which \(\phi(t)\to\phi_{\min}\) and
\(\dot\phi\to0\) as \(t\to-\infty\), lies far to the past of the plotted
interval. Thus the field sits arbitrarily close to the stabilized minimum in
the infinite past. As the universe contracts, the Klein--Gordon term
\(3H\dot\phi\) acts as antifriction because \(H<0\), driving the field away
from the minimum and into the bounce-supporting and inflationary region of the
potential. At the bounce, \(\dot\phi=0\), and the scalar reaches a
turning point. After the bounce, \(H>0\), the same term becomes ordinary
Hubble friction, and the field rolls back toward the minimum on the expanding
branch. Subsequently, the field oscillates about that minimum and eventually approaches
the late-time de Sitter asymptotics. Reheating is assumed to occur in the conventional manner at the end of
inflation \cite{Kofman:1994rk,Kofman:1997yn,Shtanov:1994ce}, and we do not
address the details of that process here.

During the entire eternal evolution, the stress tensor and curvature invariants
remain finite and all relevant energy scales stay within the regime of
control of the effective field theory. The bounce occurs at finite nonzero
radius, the inflationary scale is sub-Planckian, and the solved branch never
approaches a curvature singularity.

\subsection{Regularity and completeness}

The background is everywhere regular. The scale factor remains strictly positive,
the Hubble parameter is finite, and the curvature invariants remain bounded
throughout the evolution. The spacetime is geodesically complete: the
contracting branch extends to the infinite past, the expanding branch extends to
the infinite future, and there is no curvature singularity, coordinate
pathology, or boundary at finite affine parameter.~\footnote{This explicit closed-FRW bounce solution is not in tension with the
Borde--Guth--Vilenkin theorem \cite{Borde:2001nh}, whose past-incompleteness conclusion assumes a
positive past-directed averaged expansion condition along the relevant
geodesics; that hypothesis is not satisfied on the complete geodesics of the
present solution.} A direct asymptotic
completeness check for the explicit continued homogeneous solution is given
in App.~\ref{app:direct_completeness}.

The total expansion from the bounce to the end of inflation is large,
\begin{equation}
 N_{\rm tot}\simeq 98.68,
\label{eq:Ntot}
\end{equation}
so the standard CMB pivot window lies comfortably within the solved expanding
branch. For the benchmark choices \(N_*=55\) and \(N_*=60\), the corresponding
pivot times occur long after the bounce:
\[
\begin{aligned}
N_{\rm tot}-N_* &\simeq 43.7 \qquad (N_*=55),\\
N_{\rm tot}-N_* &\simeq 38.7 \qquad (N_*=60).
\end{aligned}
\]
The directly bounce-sensitive modes are therefore pushed to the extreme
infrared, whereas the observable pivot modes are generated on the later
curvature-diluted slow-roll branch. This is why the direct low-\(n\) evolution
is used as a regularity diagnostic, while the observable pivot predictions are
computed using the slow-roll analysis of Sec.~\ref{sec:observables}. The role
of the bounce is to complete the spacetime and provide a regular origin for the
inflationary phase; the role of the plateau is to generate the observed nearly
scale-invariant perturbations.

For the explicit tuned model, the post-inflationary scalar undergoes a long
oscillatory matter-like transient around the strictly positive minimum of the
full implemented potential. The remote past and future are asymptotically de
Sitter.

\subsection{EFT control and sub-Planckian scales}
\label{sec:eft_control}

A central requirement of the present construction is that the entire solved
branch remain within the regime of validity of ordinary low-energy effective
field theory. That is indeed the case. At the bounce one has $H=0$, so the
closed-FRW Friedmann equation gives
\begin{equation}
 \rho_{{\rm tot},b}=\frac{3}{a_b^2}.
\label{eq:bounce_density}
\end{equation}
For the explicit solution studied here,
\begin{equation}
 a_b = 7.25\times 10^4,
 \qquad
 \rho_{{\rm tot},b}\simeq 5.70\times 10^{-10}\,\mpl^4,
\label{eq:bounce_density_num}
\end{equation}
which is comfortably sub-Planckian. The inflationary pivot scales are likewise
well below the Planck scale, with $H_*\sim 7\times 10^{-6}\mpl$ and
$V_*^{1/4}\sim 3.4\times 10^{-3}\mpl$ at the observable pivots.\footnote{Importantly, the curvature-supported completion mechanism is not intrinsically tied to the
inflationary energy scale used in the explicit model. A lower-scale
realization would require a correspondingly flatter slow-roll region in order
to keep \(A_s\simeq H_*^2/(8\pi^2\epsilon_1)\) fixed, and would therefore
predict a much smaller tensor-to-scalar ratio. The same geometric completion
mechanism is therefore compatible, in principle, with low-scale variants.
}
Thus the bounce, the inflationary phase, and the transition between them all occur within
a regime where the GR plus canonical-scalar effective description remains under
control.

To make the regular and sub-Planckian character of the solution more concrete,
Table~\ref{tab:background_invariants} collects representative energy and
curvature scales at the bounce and at the observable pivots.
\begin{table*}[t]
\caption{Representative energy and curvature scales for the explicit
curvature-bounce background, expressed in reduced Planck units
($8\pi G = 1$). All entries remain sub-Planckian at the
bounce and at the observable slow-roll pivots.}
\label{tab:background_invariants}
\begin{ruledtabular}
\begin{tabular}{lccccc}
Location & $\rho_{\rm tot}$ & $V^{1/4}$ & $R$ &
$\sqrt{R_{\mu\nu}R^{\mu\nu}}$ & $\sqrt{K}$ \\
\hline
Bounce   & $5.70\times 10^{-10}$ & $4.89\times 10^{-3}$ & $2.28\times 10^{-9}$ & $1.14\times 10^{-9}$ & $9.31\times 10^{-10}$ \\
$N_*=60$ & $1.39\times 10^{-10}$ & $3.43\times 10^{-3}$ & $5.55\times 10^{-10}$ & $2.77\times 10^{-10}$ & $2.27\times 10^{-10}$ \\
$N_*=55$ & $1.38\times 10^{-10}$ & $3.43\times 10^{-3}$ & $5.53\times 10^{-10}$ & $2.77\times 10^{-10}$ & $2.26\times 10^{-10}$ \\
\end{tabular}
\end{ruledtabular}
\end{table*}
The Table shows that, from the bounce to the
observable slow-roll pivots, both the energy density and the basic curvature
invariants, including the Kretschmann scalar,
\(K = R_{\mu\nu\lambda\sigma}R^{\mu\nu\lambda\sigma}\), remain many orders of
magnitude below unity in Planck units.

\section{Inflationary observables}
\label{sec:observables}

Because the pivot scales are generated on the slow-roll branch, the observable
predictions are controlled by the usual Hubble slow-roll hierarchy. In this section \(N_*=55\) and \(N_*=60\) are used as standard benchmark
e-fold values; rather than derived from a specific reheating history or
pivot-scale matching prescription.
\begin{equation}
 \epsilon_1\equiv -\frac{\dot H}{H^2},
 \qquad
 \epsilon_2\equiv \frac{\dot\epsilon_1}{H\epsilon_1},
\label{eq:slowroll_defs}
\end{equation}
evaluated on the exact reconstructed background. To first order,
\begin{equation}
 n_s \simeq 1-2\epsilon_1-\epsilon_2,
 \qquad
 r\simeq 16\epsilon_1.
\label{eq:nsr_defs}
\end{equation}
In the amplitude-normalized solution studied here, the background yields
\begin{align}
 N_*=55:&\qquad
 n_s = 0.96170,\qquad r = 0.00446,
\label{eq:obs55}\\
 N_*=60:&\qquad
 n_s = 0.96499,\qquad r = 0.00372.
\label{eq:obs60}
\end{align}
The corresponding Hubble and energy scales are
\begin{align}
N_*=55:&\quad H_* = 6.792\times10^{-6},\nonumber\\
&\quad V_*^{1/4}=3.430\times10^{-3},
\label{eq:H55}\\[2pt]
N_*=60:&\quad H_* = 6.801\times10^{-6},\nonumber\\
&\quad V_*^{1/4}=3.432\times10^{-3}.
\label{eq:H60}
\end{align}
Using the same first-order slow-roll amplitude estimate
\begin{equation}
A_s \simeq \frac{H_*^2}{8\pi^2\epsilon_1},
\end{equation}
the amplitude-normalized branch gives
\begin{equation}
A_s \simeq
\begin{cases}
2.10\times 10^{-9}, & N_*=55,\\[3pt]
2.52\times 10^{-9}, & N_*=60.
\end{cases}
\end{equation}
The observable pivots lie deep in the curvature-diluted slow-roll regime. On
the solved background one finds
\begin{equation}
|\Omega_K|_* \equiv \frac{1}{a_*^2H_*^2}
=
\begin{cases}
9.89\times 10^{-34}, & N_*=60,\\[3pt]
4.46\times 10^{-38}, & N_*=55.
\end{cases}
\end{equation}
Thus the CMB pivot observables are controlled by the effectively flat slow-roll
branch rather than by the bounce-sensitive infrared regime.

These values lie comfortably in the observationally viable range
\cite{Planck:2018vyg}. The scalar tilt is red, the tensor-to-scalar ratio is
small but non-negligible, and the inflationary scale remains well below the
Planck scale near the GUT range. The completion and the observable phase therefore belong to one
and the same closed-FRW cosmology; no additional sector is introduced to repair
the initial singularity problem after the fact.

The bounce is a global statement about the spacetime, whereas the observables
are a local statement about the late inflationary plateau. Current observations probe
only the latter. Many nearby global potentials can therefore share the same
observed \((n_s,r)\) while preserving the same geometric mechanism for
geodesic completeness.

\section{Linear perturbations and infrared regularity}
\label{sec:perturbations}

The perturbation analysis has two distinct purposes. First, the observable pivot-scale predictions are generated on
the later curvature-diluted slow-roll branch and are well captured by the
standard slow-roll analysis of Sec.~\ref{sec:observables}. Second, direct
evolution of the closed-universe perturbations probes the infrared sector most
sensitive to the bounce itself. The latter calculation should be read
as a test of linear regularity of the background through the
bounce and inflationary era, rather than as a brute-force derivation of the
observable CMB spectrum from astronomical closed-harmonic labels. 

We expand perturbations in discrete harmonics on $S^3$, labeled by an integer
$n$. In our conventions, the propagating gauge-invariant scalar sector
and the transverse-traceless tensor sector both begin at $n=3$; the lower labels
$n=0,1,2$ correspond to non-propagating background, gauge, or global modes and
are excluded \cite{Langlois:1994qz,Bonga:2016iuf,Bonga:2017tensorclosed,Kiefer:2022qno}. Throughout this section we refer to the scanned infrared regime
$3\le n\le 60$ simply as the tested low-$n$ range. For a single minimally coupled canonical scalar field on an FRW background, the
linear vector sector is nondynamical: it is fixed by the constraint equations,
opposed to sourced by an independent matter vector degree of freedom. Thus only
the scalar and tensor sectors contain propagating linear perturbations. Our analysis is formulated directly in the gauge-invariant scalar mode
$Q_n$ and in the transverse-traceless tensor sector. Accordingly, the reported
scalar regularity and inflation-era freeze of $\zeta_n$ are physical statements
about the solved background rather than artifacts of a particular scalar gauge
fixing.~\footnote{Related analyses of curved-universe primordial spectra and quantum initial
conditions include Refs.~\cite{Handley:2019anl,Letey:2022hdp}.}

\subsection{Tensor sector}

Writing the tensor perturbation amplitude as $h_n(t)$, the physical cosmic-time
evolution equation is
\begin{equation}
 \ddot h_n + 3H\dot h_n + \frac{n^2-1}{a^2 \rc^2}\,h_n = 0,
\label{eq:tensor_mode_eq_main}
\end{equation}
where $\rc$ is the curvature radius. Equivalently, in conformal time one may
introduce the canonical variable $\mu_n=a h_n$, obeying
\begin{equation}
 \mu_n''+\left(\frac{n^2-1}{\rc^2}-\frac{a''}{a}\right)\mu_n=0.
\label{eq:tensor_canonical_eq_main}
\end{equation}
This canonical form is convenient for identifying adiabatic initial conditions on
the contracting branch, while the cosmic-time form
Eq.~\eqref{eq:tensor_mode_eq_main} is numerically more robust through
turning-point regions where the effective frequency changes sign.

We evolved the tensor modes over the tested infrared range on the explicit
background. Every mode propagated smoothly through the bounce and the
subsequent inflationary phase. No mode exhibited a breakdown of the linear
evolution or an ambiguity in the regular evolution variable. The effective
frequency generically undergoes a sign change during the evolution, as is
expected for infrared closed-universe modes, but these turning points do not
signal any pathology of the background. They are handled regularly by the
physical evolution of \(h_n(t)\).

The cutoff of the direct scan does not identify the observable CMB
window. It is a deliberately infrared diagnostic range chosen to probe the
modes most sensitive to the bounce. For sufficiently large \(n\), the
closed-harmonic eigenvalue term dominates the effective conformal-time
frequency through the bounce region, so such modes remain adiabatic there and
are not expected to provide the most stringent bounce-regularity test. The
observable pivots correspond to much higher closed-harmonic labels on the
solved background and are treated through the slow-roll analysis of
Sec.~\ref{sec:observables}. A diagnostic comparison of the relevant effective-frequency terms is given in
Appendix~\ref{app:frequency-hierarchy}.

A modest diagnostic continuation beyond the end of inflation shows that
representative tensor modes freeze as expected after horizon exit. Thus the
tensor modes exhibit the behavior one wants from a regular curvature-bounce
inflationary background: smooth propagation through the bounce, no loss of
control in the infrared, and the standard late-time approach to constant
superhorizon amplitude.

\subsection{Scalar sector}

The scalar modes are more delicate and therefore more informative. We evolve the regular gauge-invariant closed-universe scalar
mode $Q_n(t)$ directly in cosmic time. The canonicalized variable
\begin{equation}
 v_n(\eta)=\frac{a\,Q_n}{\sqrt{\Ucal_n}}
 \equiv s_n Q_n,
 \qquad
 s_n\equiv \frac{a}{\sqrt{\Ucal_n}},
\label{eq:vn_main}
\end{equation}
is used only for adiabatic initialization and frequency diagnostics. The
canonicalization factor is
\begin{equation}
 \Ucal_n
 =
 1+\frac{3}{n^2-4}\left(\frac{\dot\phi}{H}\right)^2_{\rm reg},
\label{eq:Un_main}
\end{equation}
where the subscript ``reg'' denotes the numerical regularization used to handle
the removable singular representation near $H=0$ on the solved background grid.
The physically relevant curvature perturbation on the inflationary branch is
then reconstructed as
\begin{equation}
 \zeta_n=\left(\frac{H}{\dot\phi}\right)_{\rm reg}Q_n.
\label{eq:zeta_main}
\end{equation}

The evolved variable obeys
\begin{equation}
 \ddot Q_n + b_n(t)\,\dot Q_n + c_n(t)\,Q_n = 0,
\label{eq:Qn_main}
\end{equation}
with
\begin{equation}
 b_n(t)=3H-\frac{\dot{\Ucal}_n}{\Ucal_n}.
\label{eq:bn_main}
\end{equation}
The exact mode-dependent coefficient $c_n(t)$, together with the reconstructed
background quantities entering it, is given in
App.~\ref{app:scalar_system}. For adiabatic initialization we use the
effective conformal-time equation
\begin{equation}
 v_n''+\omega_n^2(\eta)\,v_n=0,
\label{eq:vn_eff_main}
\end{equation}
where $\omega_n^2$ is likewise defined in App.~\ref{app:scalar_system}.
Adiabatic initial data are imposed only on the contracting branch, and only when
\begin{equation}
 \omega_n^2>0,
 \qquad
 \left|\frac{\omega_n'}{\omega_n^2}\right|<\epsilon_{\rm ad},
 \qquad
 \Ucal_n>0,
\label{eq:adiabatic_main}
\end{equation}
with $\epsilon_{\rm ad}=5\times10^{-2}$ by default and a minimum admissible
window of three grid points. The initialization point is chosen to be the
earliest sample in the first valid contracting-branch window. The detailed
implementation is recorded in App.~\ref{app:scalar_system}
\cite{Langlois:1994qz,Bonga:2016iuf,Bonga:2017tensorclosed,Kiefer:2022qno}.

The scalar analysis is performed over the same tested infrared range. All
scalar modes propagated regularly through the bounce and the inflationary era.
There were no ambiguous modes, no failed mode evolutions, and no pathologies in
the canonicalization factor $\Ucal_n$: for all tested modes,
$\Ucal_n$ remained positive and finite. This is a nontrivial result, since the
infrared scalar sector is the place where a hidden instability would most
naturally appear.

For our explicit solved background, the bounce occurs at
$H_b=\dot\phi_b=0$. A local analytic expansion of the symmetric closed-FRW
bounce, recorded in App.~\ref{app:local_bounce_regular}, shows that both $H$
and $\dot\phi$ vanish linearly in $x=t-t_b$ at leading order. As a result,
$\dot\phi/H$ approaches a finite continuum limit at the bounce rather than
exhibiting a true divergence. On the solved background that limit is numerically
very close (indistinguishable) to unity. Consequently the closed-universe scalar canonicalization
factor $\Ucal_n$ has a genuine finite bounce limit in the tested infrared
sector, the first-derivative coefficient $b_n(t)$ vanishes linearly at the
bounce, and the scalar mode equation has an ordinary point there. The apparent
singular structures in $\Ucal_n$ and in the coefficient $c_n(t)$ are therefore
removable in the continuum theory, with the implemented interpolation
reconstructing finite bounce values rather than masking a true divergence.

\begin{figure}[t]
\safeincludegraphics[width=\columnwidth]{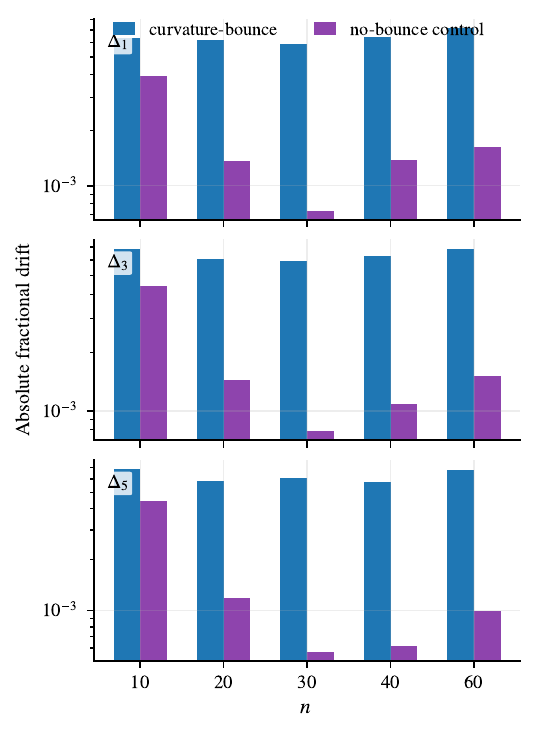}
\caption[Inflation-era freeze of the physical curvature perturbation $\zeta_n$.]{
Inflation-era freeze of the physical curvature perturbation $\zeta_n$.
The bars show
\(\Delta_k(n)=\left||\zeta_n(t_{\rm end})|/
|\zeta_n(t_{N_{\rm end}-k})|-1\right|\)
for \(k=1,3,5\), where \(t_{\rm end}=t_{\epsilon_1=1}\).
Blue bars denote the curvature-bounce background and purple bars denote the
no-bounce control. For representative closed-universe modes
\(n=10,20,30,40,60\), small values of \(\Delta_k\) indicate that
\(|\zeta_n|\) is approaching a constant on superhorizon scales. The
curvature-bounce model and the no-bounce control both exhibit qualitative
freeze-out behavior, although in the tested infrared range the curvature-bounce
branch freezes somewhat more slowly.
}
\label{fig:zeta_freeze}
\end{figure}

A dedicated inflation-era freeze check shows that the physical curvature
perturbation behaves as expected. For the representative modes shown in
Fig.~\ref{fig:zeta_freeze}, $n=10,20,30,40,60$, the fractional drift in
$|\zeta_n|$ over the last $1$, $3$, and $5$ e-folds before $\epsilon_1=1$
remains small, with the last-$5$-e-fold drift at the few-$\times 10^{-3}$
level. Thus the physical curvature perturbation freezes during inflation in the
same qualitative sense as in the corresponding no-bounce control, although in
the tested infrared range the curvature-bounce branch freezes somewhat more
slowly. The relevant agreement is qualitative rather than mode-by-mode in absolute amplitude: the
curvature-bounce background can shift the infrared normalization of the lowest
modes, while the common approach to a constant late-time amplitude is the
physically relevant inflationary test. We define
\begin{equation}
\Delta_k(n)\equiv
\left|
\frac{|\zeta_n(t_{\rm end})|}{|\zeta_n(t_{N_{\rm end}-k})|}-1
\right|,
\qquad k\in\{1,3,5\},
\label{eq:Delta_k_def}
\end{equation}
with $t_{\rm end}=t_{\epsilon_1=1}$ in the present comparison.

We also checked that these scalar-sector conclusions are stable under
reasonable variations of the interpolation threshold, the adiabatic-window
criterion, and the initialization-point choice within the first admissible WKB
window. Across these tests, no mode failures or $\Ucal_n$ pathologies appeared
in the representative set, regular bounce passage was preserved, and the
inflation-era behavior of $|\zeta_n|$ was qualitatively unchanged. Additional
deeper-infrared tests show a stable residual-drift pattern rather than exact
uniform freeze: by the end of inflation, $n=5$ remains at the two-percent level
and $n=3$ at the several-percent level, while the higher representative modes
are already frozen at the few-$\times 10^{-3}$ level.

\begin{figure}[t]
\centering
\includegraphics[width=\columnwidth]{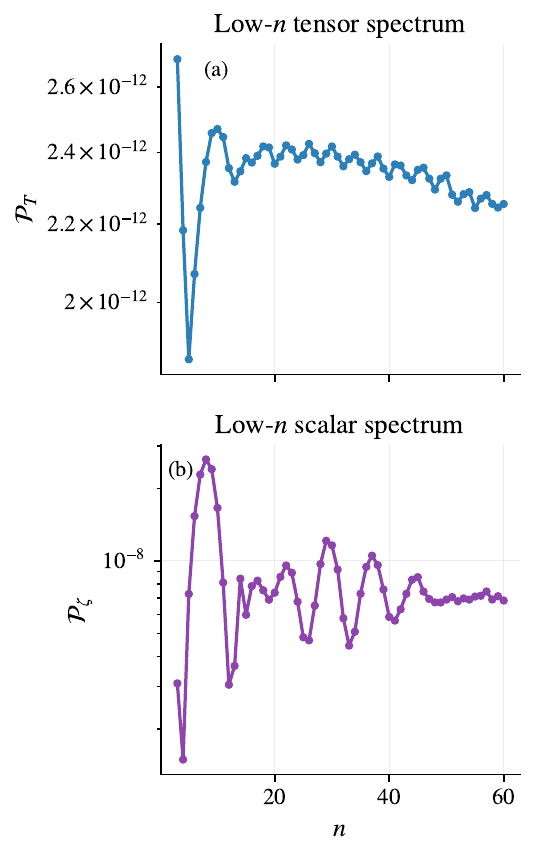}
\caption{Infrared tensor and scalar spectra for the exact closed-universe
background. Panel (a) shows the low-$n$ tensor spectrum and panel (b) the
low-$n$ scalar spectrum. In both cases the directly evolved modes probe the
bounce-sensitive infrared sector and should be interpreted as regularity
diagnostics of the curvature-bounce geometry rather than as the observable CMB
pivot spectrum. The absence of numerical pathologies in either case is part
of the evidence for linear regularity through the bounce and inflationary era.}
\label{fig:low_n_tensor_scalar_spectra}
\end{figure}

Auxiliary variables such as \(Q_n\) or the canonical scalar mode can continue to
evolve after the turning point, but that behavior is not associated to an
inflationary instability. The correct inflationary diagnostic is \(\zeta_n\),
which freezes properly during inflation on the solved background.
The broader infrared tensor and scalar diagnostics
are shown in Fig.~\ref{fig:low_n_tensor_scalar_spectra}.

\subsection{What the perturbation analysis establishes}

The perturbation analysis supports a precise and appropriately limited claim.
For the tested infrared modes, both tensor and scalar perturbations evolve
linearly and regularly through the bounce and inflationary era on the exact
background. There are no failed low-$n$ tensor or scalar modes, no ambiguous
initial-state modes, and no scalar $\Ucal_n$ pathology. Perhaps this should come as no surprise given the regularity of the bounce itself. The physical scalar
curvature perturbation freezes during inflation in the expected way, while the
tensor sector shows the expected late-time freeze after inflation.

The direct low-$n$ mode evolution should therefore be interpreted as an
infrared regularity test of the curvature-bounce background, while the
observable pivot-scale predictions are controlled by the later slow-roll branch
discussed in Sec.~\ref{sec:observables}. The two analyses are complementary:
the slow-roll calculation establishes observational viability, while the direct
infrared evolution shows that the nonsingular background remains linearly
well behaved in the regime most sensitive to the bounce.

A separate scalar-only post-inflation continuation can show late-time growth in
auxiliary scalar variables in the absence of a realistic reheating completion.
That is not part of the core inflationary claim. The robust result
needed here is stronger: on the exact solved background, the
closed-universe infrared tensor and scalar sectors evolve linearly and regularly
through the bounce and inflationary era, the scalar canonicalization factor
$\Ucal_n$ has a finite bounce limit, the continuum coefficient $c_n(t)$ has a
removable single-point bounce singularity, and the physical curvature
perturbation behaves properly during inflation itself.


\section{Discussion}
\label{sec:discussion}

Once geodesic completeness and ANEC compatibility are imposed, closed
FRW is singled out as the natural non-static nonsingular branch, and that branch
already admits an explicit realization in ordinary general relativity with a
single canonical scalar and a positive vacuum offset. The bounce is carried by
curvature, not by exotic matter. The matter sector respects the NEC and ANEC
throughout and violates only the SEC. 

Much effort has gone into constructing increasingly elaborate early-universe
models in order to evade singularity intuition. The present result shows that,
once the completeness question is posed correctly, a smooth and observationally
viable completion already exists within closed FRW in ordinary general
relativity. A common strategy is to enlarge the theory space until a
nonsingular solution appears: modified gravity, Horndeski or Galileon sectors,
higher derivatives, noncanonical kinetic terms, nonminimal couplings, or
multiple fields are introduced in order to evade the usual no-go intuition
\cite{Khoury:2001wf,Novello:2008ra,Easson:2011zy,Cai:2012va,Battefeld:2014uga,Brandenberger:2016vhg,Creminelli:2016zwa,Cai:2016thi,Cai:2017tku,Ijjas:2016tpn,Ijjas:2016vtq}.
Here the logic runs in the opposite direction. The completeness results first
identify the selected FRW branch, and only then does one ask whether minimal
physics suffices. This paper shows that it does.

The same viewpoint fixes the status of the potential. The potential used here is
not a unique law, extracted from data. Observations probe only a narrow segment
of the late inflationary plateau, so many nearby global profiles can reproduce
essentially the same observable \((n_s,r)\) while preserving the same
curvature-supported completion of the background. What matters is the existence of a non-unique reconstructed class with the
required qualitative structure: bounce region, inflationary plateau, graceful
exit, and stable minimum. The present model should therefore be read as a
concrete representative of a broader closed-universe canonical-scalar class. An
important next step would be to understand whether this qualitative structure
can arise naturally from a more fundamental particle-physics or UV-complete
setting, rather than only as a reconstructed effective description.\footnote{
Because the bounce in the present construction is supplied by positive spatial
curvature rather than by exotic scalar dynamics, any possible UV embedding need
not generate the bounce itself. It would only need to produce an
effective single-field potential with a flat inflationary region, graceful
exit, and a stabilized minimum---a qualitative structure not unfamiliar from
moduli- or axion-based constructions. A fully controlled embedding with a
positive vacuum offset, however, remains an open question.}

Phenomenologically, the model is viable in the standard inflationary sense. The
observable window lies on the later slow-roll branch, where curvature has
already been exponentially diluted and the usual inflationary observables are
well behaved and sub-Planckian. The direct perturbation analysis establishes
something complementary: the exact nonsingular background remains linearly under
control in the infrared regime most sensitive to the bounce. Tensor modes pass
smoothly through the turning-point region and freeze after inflation as
expected. Scalar modes remain regular, the closed-universe canonicalization
factor stays finite and positive, and the physical curvature perturbation
\(\zeta_n\) freezes during inflation in the standard way.

It is also important to distinguish the present construction from scenarios in
which the observed perturbations are generated or strongly processed by the
bounce itself. In some classical bouncing cosmologies, the bounce can generate
large non-Gaussianities, potentially challenging perturbativity and
observational bounds~\cite{Gao:2014eaa}. In the present model, the observable
pivot modes are generated on the later curvature-diluted slow-roll branch, long
after the bounce, while the directly evolved low-\(n\) modes are used as
infrared regularity diagnostics. A dedicated bispectrum calculation for the
extreme-infrared sector would be interesting, but is not required for the
pivot-scale viability claim made here.

Hence, what we have
constructed is an explicit homogeneous closed-FRW solution together with a
direct linear-regularity analysis on the solved background. A compact
homogeneous shear estimate shows that sufficiently small \(a^{-6}\) shear stays
subdominant on the explicit branch. We also performed a first-pass check of
small spatial-gradient effects using the existing closed-universe scalar-mode
pipeline: in the adopted harmonic normalization, the near-bounce gradient and
kinetic backreaction proxies built from the evolved \(Q_n\) modes remain small,
including under a simple full-multiplet degeneracy estimate on \(S^3\). These
checks support the robustness of the homogeneous solution, but are clearly not a
substitute for a full second-order gauge-invariant backreaction calculation or
a nonlinear inhomogeneous evolution. A complete treatment of generic
anisotropic or inhomogeneous departures from the closed-FRW branch is left for
future work.

This construction should also not be read as a measure-theoretic statement
about generic initial data in the space of closed scalar-field cosmologies.
It was argued that perpetually bouncing closed scalar-field universes can form a
special, fractal, measure-zero set in minisuperspace~\cite{Page:1984qt}. The
present work addresses a different question: whether the closed-FRW branch
selected by geodesic completeness and ANEC compatibility admits an explicit
minimal realization with viable slow-roll observables and controlled linear
infrared perturbations. A full measure or basin-of-attraction analysis is
beyond the scope of this paper.

The relevance of this model does not require a dramatic observational smoking gun. Any direct signature of the bounce is most likely to appear in the
extreme infrared, precisely where cosmic variance is most severe. The real
achievement is more structural than spectral: a smooth, complete,
observationally viable homogeneous early-universe history exists in a framework
much leaner than is usually assumed.

The conceptual simplification is that the bounce is not a creation event or a
patch onto an otherwise incomplete spacetime. It is an ordinary part of a
single geodesically complete classical history. Since the model is not cyclic, the
usual entropy accumulation problem of repeated-cycle cosmologies does not arise \cite{Tolman:1934}.
Inflation is not what obstructs completeness; for a nontrivial FRW universe, accelerated
expansion is part of the cure. Nor is the bounce by itself the main message.
The geometry instead points to a definite trio of ingredients: positive
curvature, inflation, and a positive vacuum offset, and the present
construction shows that this trio is enough to realize an explicit homogeneous
completion of the early universe within ordinary low-energy physics. If one
wants the simplest known route to a geodesically complete inflationary
cosmology, closed FRW is not a curiosity. It is the natural place to begin.

\section{Conclusion}
\label{sec:conclusion}

We have exhibited an explicit geodesically complete bounce-plus-inflation
solution in a closed \(k=+1\) FRW universe within ordinary general relativity,
sourced by a single canonical scalar field with a positive vacuum offset. The
bounce is curvature-supported rather than NEC-violating. The matter sector
satisfies the NEC (and ANEC) throughout and violates only the SEC.

The resulting background is smooth, nonsingular, sub-Planckian on the solved
branch, and contains a curvature-diluted slow-roll era with viable benchmark
values of \(n_s\) and \(r\). Direct evolution of the closed-universe infrared
scalar and tensor modes further shows that the exact background remains
linearly regular through the bounce and inflationary era in the regime most
susceptible to the bounce.

The force of the result is therefore not a special potential and not a dramatic
observable bounce signature. Rather, the ingredients selected by the
completeness analysis: positive spatial curvature, inflation,
and a positive vacuum offset realize a smooth, complete, and observationally
viable homogeneous cosmology within the simplest classical framework available.
In that sense, the model constructed here is more than a worked example. It is
a concrete demonstration that geodesic completion of the early universe can be
achieved without leaving the minimal closed-FRW setting.

\begin{acknowledgments}
DAE is supported
in part by the U.S. Department of Energy, Office of High Energy Physics, under
Award Number DE-SC0019470.
\end{acknowledgments}

\appendix

\section{Generator equations for the production branch}
\label{app:production_generator}

The exact solved background used in the production analysis is generated by the
script \path{scripts/make_versionB_background.py}. The amplitude-normalized
production branch is obtained from the benchmark family by the rescaling
\begin{align}
A &\to \sqrt{\lambda}\,A, &
V_0 &\to \lambda V_0, \nonumber\\
m^2 &\to \lambda m^2, &
\Lambda &\to \lambda \Lambda,
\label{eq:generator_family_rescaling_appendix}
\end{align}
with the production value of \(\lambda\) listed in
Table~\ref{tab:model_parameters}. All other parameters appearing below are
likewise fixed by Table~\ref{tab:model_parameters}. In the final production
branch, the reheating/minimum interpolation is implemented with a
compact-support switch so that the plateau contribution vanishes identically
for sufficiently large \(\phi\) and the late-time minimum is controlled exactly
by \(\Lambda\).

We define
\begin{equation}
z \equiv \frac{\phi-\phi_0+\Delta_L}{\Delta_L},
\label{eq:z_prod_appendix}
\end{equation}
and
\begin{equation}
f(z)=e^{-1/z},
\qquad
g(z)=e^{-1/(1-z)}.
\label{eq:fg_appendix}
\end{equation}
The smooth compact-support interpolation is
\begin{equation}
\Theta_L(\phi)=
\begin{cases}
0, & z\le 0,\\[3pt]
1, & z\ge 1,\\[3pt]
\dfrac{f(z)}{f(z)+g(z)}, & 0<z<1,
\end{cases}
\label{eq:ThetaL_appendix}
\end{equation}
with
\begin{equation}
f'(z)=\frac{f(z)}{z^2},
\qquad
g'(z)=-\,\frac{g(z)}{(1-z)^2},
\label{eq:fgprime_appendix}
\end{equation}
and
\begin{equation}
\Theta_L'(\phi)=
\begin{cases}
0, & z\le 0\ \text{or}\ z\ge 1,\\[6pt]
\dfrac{1}{\Delta_L}\,
\dfrac{f'(z)g(z)-f(z)g'(z)}{[f(z)+g(z)]^2},
& 0<z<1.
\end{cases}
\label{eq:ThetaLprime_appendix}
\end{equation}

The bounce-side potential pieces are
\begin{align}
V_{\rm quad}(\phi)
&=4A^2\!\left(5e^{-(\phi-\phi_0)}-2e^{-2(\phi-\phi_0)}\right),
\label{eq:Vquad_prod_appendix}\\
V_{\rm wall}(\phi)
&=12A^2+\frac{12A^2}{\gamma}
\left(e^{-\gamma(\phi-\phi_0)}-1\right).
\label{eq:Vwall_prod_appendix}
\end{align}

The piece \(V_{\rm quad}\) is not a phenomenological ansatz: it is the exact
canonical-scalar potential associated with the closed quadratic
curvature-bounce family producing \(a \sim (t^2+c)\), in the prototype notation 
 of \cite{Easson:2024fzn}. In the present production generator that family is
written in shifted and rescaled bounce-scale conventions, with the bounce
radius encoded by \(a_b=1/(2A)\),
so that,
\begin{equation}
V_b(\phi)=\Theta_L(\phi)\,V_{\rm quad}(\phi)
+\bigl[1-\Theta_L(\phi)\bigr]V_{\rm wall}(\phi).
\label{eq:Vb_prod_appendix}
\end{equation}
Their derivatives are
\begin{align}
V_{{\rm quad},\phi}(\phi)
&=4A^2\!\left(-5e^{-(\phi-\phi_0)}+4e^{-2(\phi-\phi_0)}\right),
\label{eq:Vquadprime_prod_appendix}\\
V_{{\rm wall},\phi}(\phi)
&=-12A^2 e^{-\gamma(\phi-\phi_0)},
\label{eq:Vwallprime_prod_appendix}
\end{align}
and
\begin{align}
V_{b,\phi}
&=\Theta_L'(\phi)\bigl(V_{\rm quad}-V_{\rm wall}\bigr)
+\Theta_L(\phi)V_{{\rm quad},\phi}
\nonumber\\
&\quad
+\bigl[1-\Theta_L(\phi)\bigr]V_{{\rm wall},\phi}.
\label{eq:Vbprime_prod_appendix}
\end{align}

The smooth switch to the inflationary sector is
\begin{align}
S_1(\phi)
&=\frac12\left[1+\tanh\!\left(\frac{\phi-\phi_{\rm trans}}{\delta_1}\right)\right],
\label{eq:S1_prod_appendix}\\
S_1'(\phi)
&=\frac{1}{2\delta_1}\,
\sech^2\!\left(\frac{\phi-\phi_{\rm trans}}{\delta_1}\right).
\label{eq:S1prime_prod_appendix}
\end{align}

The plateau and minimum pieces are
\begin{align}
V_{\rm pl}(\phi)
&=\frac{V_0}{\left[1+u(\phi)\right]^2},
\qquad
u(\phi)\equiv e^{\beta(\phi-\phi_p)},
\label{eq:Vpl_prod_appendix}\\
V_{\rm min}(\phi)
&=\frac12 m^2(\phi-\phi_m)^2,
\label{eq:Vmin_prod_appendix}
\end{align}
with derivatives
\begin{align}
V_{{\rm pl},\phi}(\phi)
&=-\,\frac{2\beta V_0\,u(\phi)}{\left[1+u(\phi)\right]^3},
\label{eq:Vplprime_prod_appendix}\\
V_{{\rm min},\phi}(\phi)
&=m^2(\phi-\phi_m).
\label{eq:Vminprime_prod_appendix}
\end{align}

The reheating/minimum transition is implemented with the same compact-support
profile as \(\Theta_L\), but centered on \(\phi_{\rm reh}\). Defining
\begin{equation}
w_{\rm reh}=2.4119\,\delta_2,
\qquad
z_{\rm reh}\equiv \frac{\phi-\phi_{\rm reh}+w_{\rm reh}}{2w_{\rm reh}},
\label{eq:zreh_prod_appendix}
\end{equation}
one has
\begin{equation}
T(\phi)=
\begin{cases}
0, & z_{\rm reh}\le 0,\\[3pt]
1, & z_{\rm reh}\ge 1,\\[3pt]
\dfrac{f(z_{\rm reh})}{f(z_{\rm reh})+g(z_{\rm reh})},
& 0<z_{\rm reh}<1,
\end{cases}
\label{eq:T_prod_appendix}
\end{equation}
and
\begin{equation}
T'(\phi)=
\begin{cases}
0, & z_{\rm reh}\notin(0,1),\\[6pt]
\displaystyle
\frac{f'g-fg'}{2w_{\rm reh}(f+g)^2},
& 0<z_{\rm reh}<1,
\end{cases}
\label{eq:Tprime_prod_appendix}
\end{equation}
where in the second line \(f,f',g,g'\) are evaluated at \(z=z_{\rm reh}\).

Equivalently, \(T(\phi)=0\) for \(\phi\le \phi_{\rm reh}-w_{\rm reh}\) and
\(T(\phi)=1\) for \(\phi\ge \phi_{\rm reh}+w_{\rm reh}\), so on the far right
\(V_{\rm inf}(\phi)=V_{\rm min}(\phi)\) exactly and the residual plateau
contribution vanishes identically.
The inflationary-sector potential is
\begin{equation}
V_{\rm inf}(\phi)=\bigl[1-T(\phi)\bigr]V_{\rm pl}(\phi)
+T(\phi)V_{\rm min}(\phi),
\label{eq:Vinf_prod_appendix}
\end{equation}
with derivative
\begin{align}
V_{{\rm inf},\phi}
&=T'(\phi)\bigl(V_{\rm min}-V_{\rm pl}\bigr)
+\bigl[1-T(\phi)\bigr]V_{{\rm pl},\phi}
\nonumber\\
&\quad
+T(\phi)V_{{\rm min},\phi}.
\label{eq:Vinfprime_prod_appendix}
\end{align}

The full production generator is therefore

\begin{equation}
V_{\rm total}(\phi)=\bigl[1-S_1(\phi)\bigr]V_b(\phi)
+S_1(\phi)V_{\rm inf}(\phi)+\Lambda,
\label{eq:Vtotal_prod_appendix}
\end{equation}

with derivative
\begin{align}
V_{{\rm total},\phi}
&=S_1'(\phi)\bigl(V_{\rm inf}-V_b\bigr)
+\bigl[1-S_1(\phi)\bigr]V_{b,\phi}
\nonumber\\
&\quad
+S_1(\phi)V_{{\rm inf},\phi}.
\label{eq:Vtotalprime_prod_appendix}
\end{align}

The parameter meanings are as follows. The parameter \(A\) sets the bounce
scale through \(a_b=1/(2A)\) and fixes the overall scale of the left bounce
sector. The parameter \(\phi_0\) is the field value used at the bounce and in
the left-sector interpolation, while \(\Delta_L\) and \(\gamma\) control the
width and monotonic stabilization of the left wall. The pair
\((\phi_{\rm trans},\delta_1)\) controls the smooth switch from the bounce
sector to the inflationary sector. The parameters
\((V_0,\beta,\phi_p)\) specify the plateau piece, with
\(\beta=\sqrt{2/3}\) fixed in the code. The pair
\((\phi_{\rm reh},\delta_2)\) sets the center and nominal width scale of the
compact-support transition from the plateau to the stabilized quadratic
minimum, with implemented half-width \(w_{\rm reh}=2.4119\,\delta_2\). The
location and curvature of that minimum are set by \((\phi_m,m^2)\). Finally,
\(\Lambda\) is the positive vacuum offset of the full potential.

The background arrays are generated by imposing the bounce initial data
\begin{align}
a(0)&=a_b=\frac{1}{2A}, &
H(0)&=0, \nonumber\\
\phi(0)&=\phi_0, &
\dot\phi(0)&=0,
\label{eq:production_bounce_initial_data_appendix}
\end{align}
and integrating the closed-FRW Einstein--scalar system
\begin{align}
\dot a &= aH,
\label{eq:prod_rhs_a_appendix}\\
\dot H &= -\frac12\dot\phi^2+\frac{1}{a^2},
\label{eq:prod_rhs_H_appendix}\\
\ddot\phi &= -3H\dot\phi - V_{{\rm total},\phi}(\phi),
\label{eq:prod_rhs_phi_appendix}
\end{align}
backward and forward in cosmic time.

In the final production run the script \path{scripts/make_versionB_background.py}
is called with the branch stem \texttt{versionB\_background\_ampnorm\_compactswitch},
the production value of \(\lambda\), and the reheating/minimum switch mode set
to \texttt{compact-support}. The mode equations were integrated using \texttt{scipy.integrate.solve\_ivp}
with relative tolerance \(\texttt{rtol}=10^{-7}\), absolute tolerance
\(\texttt{atol}=10^{-9}\), and maximum step size
\(\texttt{max\_step}=1.8\times 10^{3}\).

The bounce time is fixed by construction at \(t_b=0\). The end of inflation is
located by the terminal event
\begin{equation}
H^2+\dot H = 0,
\label{eq:prod_eps1_event_appendix}
\end{equation}
which is equivalent to \(\epsilon_1=1\) away from the bounce since
\(H^2+\dot H=-(\epsilon_1-1)H^2\). After the backward and forward branches are
combined, the saved conformal time is constructed as
\begin{equation}
\eta(t)=-\int_t^{t_{\rm end}}\frac{d\tau}{a(\tau)},
\label{eq:prod_eta_appendix}
\end{equation}
so that the stored production arrays are
\[
\{t,\eta,a,H,\phi,\dot\phi\}.
\]

\section{Closed-universe scalar perturbations and numerical implementation}
\label{app:scalar_system}

In this Appendix we record the scalar perturbation system used in the numerical
analysis. The formulas given here are faithful to the implemented pipeline. They
are not a separate analytic approximation.

\subsection{Conventions and reconstructed background quantities}
\label{app:scalar_conventions}

We work in the closed $k=+1$ FRW background
\begin{equation}
 ds^2=-dt^2+a(t)^2\,d\Omega_3^2,
\end{equation}
with
\begin{equation}
8\pi G=1,\qquad
H\equiv \frac{\dot a}{a},\qquad
\Hc \equiv \frac{a'}{a}=aH,
\label{eq:background_hubble_defs_appendix}
\end{equation}
and scalar harmonics labeled by integers
\begin{equation}
 n\ge 3.
\end{equation}
The closed-universe wavenumber reported in the numerics is
\begin{equation}
 q_n=\frac{\sqrt{n^2-1}}{\rc},
\label{eq:qn_appendix}
\end{equation}
with curvature radius $\rc=1$ by default.

Here, \(V\), \(V_{,\phi}\), and
\(V_{,\phi\phi}\) below denote the numerically nondimensionalized potential and
its derivatives, not quantities with un-restored explicit mass scales.

The perturbation solver treats the saved background arrays
\[
 \{t,\eta,a,H,\phi,\dot\phi\}
\]
as the authoritative input and reconstructs the total potential directly from
the solved background,
\begin{equation}
 \Vpot
 =
 3\left(H^2+\frac{1}{a^2}\right)-\frac{1}{2}\dot\phi^2.
\label{eq:Vtot_reconstructed}
\end{equation}
If the model is written in the form
\begin{equation}
 \Vpot(\phi)=U(\phi)+\Lambda,
\end{equation}
then Eq.~\eqref{eq:Vtot_reconstructed} is the quantity used internally by the
pipeline. The corresponding derivatives are reconstructed as
\begin{align}
 V_{,\phi} &= -\ddot\phi - 3H\dot\phi,
\label{eq:Vphi_reconstructed}\\
 V_{,\phi\phi} &= \frac{dV_{,\phi}/dt}{\dot\phi}.
\label{eq:Vphiphi_reconstructed}
\end{align}
Equation~\eqref{eq:Vphiphi_reconstructed} is an exact identity on monotonic
branches where $\dot\phi\neq 0$. In the numerical implementation,
$V_{,\phi\phi}$ is reconstructed on those branches, and isolated non-finite
samples near zeros of $\dot\phi$ are filled by local interpolation on the
solved background grid.

\subsection{Evolved scalar variable}
\label{app:evolved_Q}

The main evolved variable is the closed-universe gauge-invariant scalar mode
\begin{equation}
 Q_n(t).
\end{equation}
This is the variable integrated directly in cosmic time. The solver evolves its
real and imaginary parts as a first-order real system equivalent to the
second-order complex equation
\begin{equation}
 \ddot Q_n + b_n(t)\,\dot Q_n + c_n(t)\,Q_n = 0.
\label{eq:Qn_main_eq}
\end{equation}

The friction coefficient is
\begin{equation}
 b_n(t)=3H-\frac{\dot{\Ucal}_n}{\Ucal_n},
\label{eq:bn_appendix}
\end{equation}
with $\dot{\Ucal}_n$ obtained by differentiating a spline fit to
$\Ucal_n(t)$.

\subsection{Canonicalized variable and effective frequency}
\label{app:canonical_variable}

The canonicalized variable used only for adiabatic initialization and frequency
diagnostics is
\begin{equation}
 v_n(\eta)=\frac{a\,Q_n}{\sqrt{\Ucal_n}}
 \equiv s_n Q_n,
 \qquad
 s_n \equiv \frac{a}{\sqrt{\Ucal_n}}.
\label{eq:vn_appendix}
\end{equation}
The canonicalization factor is
\begin{equation}
 \Ucal_n
 = 1 + \frac{3}{n^2-4}\left(\frac{\dot\phi}{H}\right)^2_{\rm reg}.
\label{eq:Un_appendix}
\end{equation}
Here the subscript ``reg'' denotes the numerical regularization used in the
implementation: the raw ratio is evaluated when the denominator exceeds a fixed
threshold, and otherwise the non-finite samples are filled by one-dimensional
interpolation on the background grid. In the present implementation the
threshold is
\begin{equation}
 \delta = 10^{-14}.
\end{equation}

The conformal-time effective equation used for WKB initialization and
diagnostics is
\begin{equation}
 v_n''+\omega_n^2(\eta)\,v_n=0,
\label{eq:vn_eff_eq}
\end{equation}
with
\begin{equation}
 \omega_n^2
 =
 a^2 c_n
 -\Hc^2
 -\Hc'
 +\Hc\,\frac{\Ucal_n'}{\Ucal_n}
 -\frac34\left(\frac{\Ucal_n'}{\Ucal_n}\right)^2
 +\frac12\frac{\Ucal_n''}{\Ucal_n}.
\label{eq:omega_eff_appendix}
\end{equation}
All conformal-time derivatives in Eq.~\eqref{eq:omega_eff_appendix} are
obtained by differentiating spline representations of the corresponding
background quantities.

For the explicit solved background used in this work, the bounce occurs at
$H_b=\dot\phi_b=0$. Writing $x\equiv t-t_b$, the local background expansion
takes the form
\begin{align}
a(t) &= a_b+\frac{x^2}{2a_b}+O(x^4), \nonumber\\
H(t) &= \frac{x}{a_b^2}+O(x^3), \nonumber\\
\dot\phi(t) &= \ddot\phi_b\,x+O(x^3).
\label{eq:local_background_appendix}
\end{align}
where $\dot H_b=1/a_b^2$ follows from the closed-FRW background equations.
Hence $\dot\phi/H$ has a finite continuum limit whenever $\ddot\phi_b$ is
finite:
\begin{equation}
\lim_{t\to t_b}\frac{\dot\phi}{H}
=
\frac{\ddot\phi_b}{\dot H_b}
=
-\,a_b^2\,V_{,\phi}(\phi_b).
\label{eq:lambda_b_appendix}
\end{equation}
For the explicit solved model used in the body of the paper, this limit is
numerically very close to unity. The important point for the present Appendix is
that it is finite on the solved background. Therefore
\begin{equation}
\Ucal_n=\Ucal_{n,b}+O(x^2),
\qquad
\Ucal_{n,b}
=
1+\frac{3}{n^2-4}
\left(
\lim_{t\to t_b}\frac{\dot\phi}{H}
\right)^2,
\label{eq:Un_finite_limit_appendix}
\end{equation}
so the canonicalization factor has a genuine finite bounce limit in the
low-$n$ sector studied in the body of the paper. This statement is
model-specific: it is a property of the solved background considered here, opposed to a
general theorem for arbitrary closed-FRW bounces.

\subsection{A simple analytic representative of the local bounce geometry}
\label{app:toy_local_bounce}

It is useful to separate what is universal about a smooth closed-FRW bounce from
what is model-dependent. Equation~\eqref{eq:local_background_appendix} shows
that every smooth symmetric bounce of the type studied here has the same
leading local geometric form,
\begin{align}
a(t) &= a_b+\frac{(t-t_b)^2}{2a_b}
      +O\!\left((t-t_b)^4\right), \nonumber\\
H(t) &= \frac{t-t_b}{a_b^2}
      +O\!\left((t-t_b)^3\right).
\label{eq:universal_local_bounce_appendix}
\end{align}
A natural exact analytic representative of this local geometry is obtained by
keeping only the leading terms,
\begin{equation}
a_{\rm toy}(t)=a_b+\frac{(t-t_b)^2}{2a_b}.
\label{eq:toy_bounce_scale_factor_appendix}
\end{equation}
This toy model is not a global solution for the full cosmology. Its
role is to provide the simplest exact background with the same leading local
bounce structure as Eq.~\eqref{eq:universal_local_bounce_appendix}.

For Eq.~\eqref{eq:toy_bounce_scale_factor_appendix} one finds
\begin{equation}
H=\frac{2x}{2a_b^2+x^2},
\qquad
\dot H=\frac{2(2a_b^2-x^2)}{(2a_b^2+x^2)^2},
\qquad
x\equiv t-t_b.
\end{equation}
Using the reconstruction formula
\begin{equation}
\dot\phi^2=2\left(\frac{1}{a^2}-\dot H\right),
\end{equation}
one obtains
\begin{equation}
\dot\phi^2=\frac{4x^2}{(2a_b^2+x^2)^2},
\end{equation}
and choosing the smooth branch gives
\begin{equation}
\dot\phi=\frac{2x}{2a_b^2+x^2}=H.
\label{eq:toy_phidot_equals_H_appendix}
\end{equation}
Hence
\begin{equation}
\lambda_b \equiv \lim_{t\to t_b}\frac{\dot\phi}{H}=1
\end{equation}
exactly in the toy model.

Integrating Eq.~\eqref{eq:toy_phidot_equals_H_appendix} yields
\begin{equation}
\phi(t)=\phi_b+\ln\!\left(1+\frac{x^2}{2a_b^2}\right).
\end{equation}
The reconstructed potential is
\begin{equation}
V(t)=3H^2+\dot H+\frac{2}{a^2}
=\frac{12a_b^2+10x^2}{(2a_b^2+x^2)^2}.
\end{equation}
Writing $\Delta\phi\equiv \phi-\phi_b$, so that
\begin{equation}
e^{\Delta\phi}=1+\frac{x^2}{2a_b^2},
\end{equation}
one obtains
\begin{equation}
V(\phi)=\frac{1}{a_b^2}\left(5e^{-\Delta\phi}-2e^{-2\Delta\phi}\right).
\label{eq:toy_potential_appendix}
\end{equation}
It follows immediately that
\begin{equation}
V_{,\phi}(\phi_b)=-\frac{1}{a_b^2},
\qquad
V_{,\phi\phi}(\phi_b)=-\frac{3}{a_b^2},
\end{equation}
and therefore
\begin{equation}
\lambda_b=1,
\qquad
\nu_b\equiv a_b^2V_{,\phi\phi}(\phi_b)=-3
\label{eq:toy_lambda_nu_appendix}
\end{equation}
exactly for the analytic representative.

Substituting the toy background into the exact coefficient
Eq.~\eqref{eq:cn_appendix} gives the closed-form result
\begin{equation}
c_n^{\rm toy}(x)=\frac{4a_b^2(n^2-5)}{(2a_b^2+x^2)^2},
\label{eq:toy_cn_exact_appendix}
\end{equation}
and therefore
\begin{equation}
c_{n,b}^{\rm toy}\equiv \lim_{x\to 0}c_n^{\rm toy}(x)
=\frac{n^2-5}{a_b^2}.
\label{eq:toy_cn_limit_appendix}
\end{equation}
Thus the especially simple value of \(c_{n,b}\) entering the scalar mode
equation, Eq.~\eqref{eq:Qn_main_eq}, is exact in the toy representative, not
merely suggested by numerical extrapolation.

The significance of this toy background is interpretive rather than universal.
Equations~\eqref{eq:toy_lambda_nu_appendix}
and~\eqref{eq:toy_cn_limit_appendix} do not provide a theorem for arbitrary
closed-FRW bounces; in general, the bounce invariants $\lambda_b$ and $\nu_b$,
and the exact value of $c_{n,b}$, remain model-dependent. What the toy model
shows is that the values $\lambda_b=1$, $\nu_b=-3$, and
$c_{n,b}=(n^2-5)/a_b^2$ arise exactly in the simplest analytic realization of
the universal local bounce geometry
Eq.~\eqref{eq:universal_local_bounce_appendix}. The fact that the solved
background used in the main text lands numerically very close to these values
should be viewed as analytically intelligible rather than accidental.

\subsection{Exact coefficient \texorpdfstring{$c_n(t)$}{c\_n(t)}}
\label{app:cn_exact}

The coefficient \(c_n(t)\) appearing in Eq.~\eqref{eq:Qn_main_eq} is
implemented from the standard closed-universe gauge-invariant scalar formalism
used in
Refs.~\cite{Langlois:1994qz,Bonga:2016iuf,Bonga:2017tensorclosed,Kiefer:2022qno},
with the background inputs \(\dot\phi\), \(V_{,\phi}\), and
\(V_{,\phi\phi}\) reconstructed from the solved branch as described above.
It is written in the form
\begin{equation}
c_n=(8\pi G)\,\frac{\mathcal N_n}{\mathcal D_n},
\label{eq:cn_appendix}
\end{equation}
with
\begin{equation}
\mathcal D_n
=
a^2\dot a^2\left[2(n^2-4)\dot a^2+8\pi G\,a^2\dot\phi^2\right],
\label{eq:Dn_appendix}
\end{equation}
and
{\scriptsize
\begin{align}
\mathcal N_n ={}& \mathcal N_n^{(1)}+\mathcal N_n^{(2)}+\mathcal N_n^{(3)}
\nonumber\\
&+\mathcal N_n^{(4)}+\mathcal N_n^{(5)},
\label{eq:Nn_appendix}
\end{align}
where
\begin{align}
\mathcal N_n^{(1)}={}&
\frac{\dot a^4 (n^2-4)\left[(n^2-1)+a^2V_{,\phi\phi}\right]}{4\pi G},
\label{eq:Nn1_appendix}\\
\mathcal N_n^{(2)}={}&
(4n^2-7)a^3\dot a^3\dot\phi\,V_{,\phi},
\label{eq:Nn2_appendix}\\
\mathcal N_n^{(3)}={}&
-\pi G\,\frac{n^2-1}{n^2-4}\,a^4\dot\phi^4
\nonumber\\
&\qquad\times\left[8\pi G\,a^2(\dot\phi^2+2\Vpot)-6\right],
\label{eq:Nn3_appendix}\\
\mathcal N_n^{(4)}={}&
(n^2-1)a^2\dot a^2
\biggl[
-6\pi G\,\frac{n^2-5}{n^2-4}\,a^2\dot\phi^4
+4\pi G\,a^2\dot\phi^2\Vpot
\nonumber\\
&\qquad +\frac32\dot\phi^2
+\frac92\dot a^2\dot\phi^2
\biggr],
\label{eq:Nn4_appendix}\\
\mathcal N_n^{(5)}={}&
a^3\dot a
\bigl[
a\dot a\,\dot\phi^2V_{,\phi\phi}
+2a\dot a\,(V_{,\phi})^2
\nonumber\\
&\qquad
+4\pi G\,a^2\dot\phi\,V_{,\phi}\,(\dot\phi^2+2\Vpot)
\nonumber\\
&\qquad -\dot\phi\,V_{,\phi}
\bigr].
\label{eq:Nn5_appendix}
\end{align}
}
Thus, at the continuum level, the apparent bounce pole in $c_n$ is removable
for the smooth background considered here. The denominator
$\mathcal D_n$ vanishes as $(t-t_b)^4$ directly from
Eq.~\eqref{eq:local_background_appendix}. For the numerator, the quartic scaling is not a matter of naive term-by-term
power counting alone. Among the numerator pieces, the only naively dangerous
contribution is \(\mathcal N_n^{(5)}\), since \(a^3\dot a=O(x)\) while the
bracket appears to contain \(O(x)\) terms. Expanding near the bounce with
\(\dot\phi=-V_{,\phi}(\phi_b)x+O(x^3)\) and \(V_b=3/a_b^2\), the leading
\(O(x)\) pieces in that bracket cancel exactly, so
\(\mathcal N_n^{(5)}=O(x^4)\). Using the local bounce expansion together with
the bounce identities \(H_b=\dot\phi_b=0\), \(V_b=3/a_b^2\), and
\(\dot H_b=1/a_b^2\), the remaining lower-order pieces cancel as well, leaving
\begin{equation}
\mathcal N_n=O\!\left((t-t_b)^4\right),
\qquad
\mathcal D_n=O\!\left((t-t_b)^4\right).
\end{equation}

Accordingly $c_n(t)$ has a finite continuum bounce limit. In the present
implementation, the exact bounce sample is not replaced by explicit analytic
limit evaluation; instead, the single undefined grid point is filled
numerically by local interpolation. For the representative modes used in the
audit, this interpolation agrees very well with local limit extrapolation,
although at the deepest tested mode $n=3$ the discrepancy remains at the
percent level. The code therefore reconstructs the finite continuum value
accurately, but it would be too strong to say that it derives that limit
analytically for the full solved background.

\subsection{Local bounce expansion and analytic regularity of the scalar sector}
\label{app:local_bounce_regular}

Combining the local expansion
Eq.~\eqref{eq:local_background_appendix} with the finite-limit result
Eq.~\eqref{eq:Un_finite_limit_appendix}, one finds
\begin{equation}
\Ucal_n=\Ucal_{n,b}+O(x^2),
\qquad
\dot{\Ucal}_n=O(x),
\end{equation}
and therefore
\begin{equation}
b_n(t)=3H-\frac{\dot{\Ucal}_n}{\Ucal_n}=O(x).
\end{equation}
Together with the removable finite bounce limit of $c_n(t)$, this gives the
local scalar mode equation
\begin{equation}
\ddot Q_n + O(x)\,\dot Q_n + \left(c_{n,b}+O(x^2)\right)Q_n=0.
\label{eq:Qn_local_regular_appendix}
\end{equation}
Hence $x=0$ is an ordinary point of the scalar perturbation equation on the
solved background. In particular, each scalar mode admits a regular Taylor
expansion through the bounce,
\begin{equation}
Q_n(x)=Q_{n,b}+\dot Q_{n,b}\,x-\frac12 c_{n,b}Q_{n,b}\,x^2+O(x^3),
\end{equation}
with freely specifiable finite data $(Q_{n,b},\dot Q_{n,b})$.

This result is local and model-specific. It does not constitute a general
theorem for arbitrary closed-FRW bounces. What it establishes is that, for the
exact solved symmetric bounce background used in this work, the scalar
perturbation equation is analytically regular at the bounce. This provides a
local continuum underpinning for the numerical infrared regularity observed in
the body of the paper.

\subsection{Adiabatic initial data}
\label{app:adiabatic_init}

Adiabatic initial data are imposed only on the contracting branch. A sample is
declared admissible only if all of the following hold:
\begin{equation}
 \omega_n^2>0,
 \qquad
 \left|\frac{\omega_n'}{\omega_n^2}\right|<\epsilon_{\rm ad},
 \qquad
 \Ucal_n>0,
 \qquad
 \Ucal_n\ {\rm finite},
\label{eq:adiabatic_criterion_appendix}
\end{equation}
with default value
\begin{equation}
 \epsilon_{\rm ad}=5\times10^{-2}.
\end{equation}
The code requires at least three contiguous admissible grid points and chooses
the earliest sample in the first valid window.

The adiabaticity parameter used in practice is
\begin{equation}
 \mathcal A_n \equiv \left|\frac{\omega_n'}{\omega_n^2}\right|,
\label{eq:adiabaticity_appendix}
\end{equation}
with $\mathcal A_n=\infty$ whenever $\omega_n^2\le 0$.

Once $\eta_{\rm init}$ is chosen, the first-order WKB data imposed on $v_n$ are
\begin{align}
 v_n(\eta_{\rm init})
 &=
 \frac{1}{\sqrt{2\omega_n(\eta_{\rm init})}},
\label{eq:wkb_v_appendix}\\
 v_n'(\eta_{\rm init})
 &=
 \left(
 -i\omega_n
 -\frac12\frac{\omega_n'}{\omega_n}
 \right)_{\eta_{\rm init}}
 v_n(\eta_{\rm init}).
\label{eq:wkb_vprime_appendix}
\end{align}
These are then converted into initial data for the evolved variable:
\begin{align}
 Q_n &= \frac{v_n}{s_n},
\label{eq:Qinit_appendix}\\
 Q_n' &= \frac{v_n' - s_n' Q_n}{s_n},
\label{eq:Qprime_init_appendix}\\
 \dot Q_n &= \frac{Q_n'}{a}.
\label{eq:Qdot_init_appendix}
\end{align}

\subsection{Effective-frequency hierarchy}
\label{app:frequency-hierarchy}

\begin{figure}[t]
\centering
\includegraphics[width=\columnwidth]{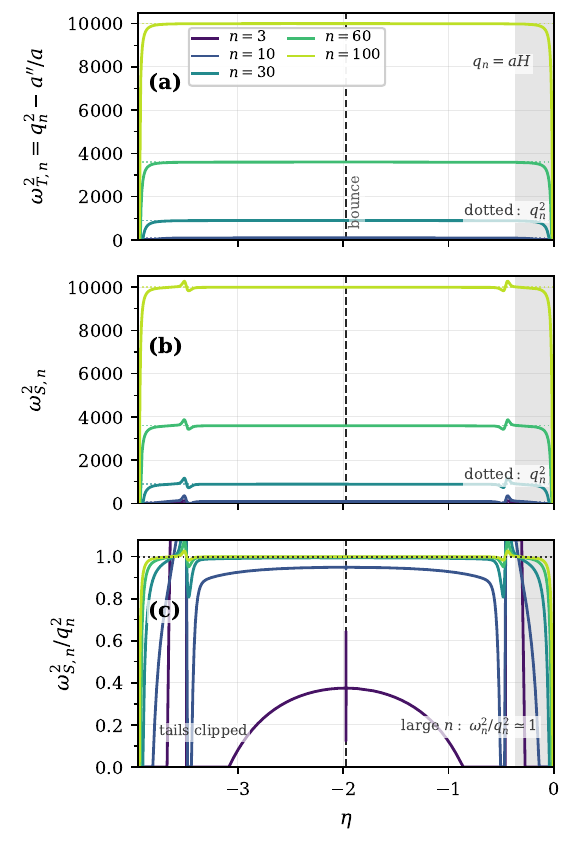}
\caption{
Effective conformal-time frequency diagnostics for representative closed
harmonics on the final compact-support branch. The dashed vertical line marks
the bounce, and the shaded band indicates the approximate range where
\(q_n=aH\) for the displayed modes. Panel (a) shows the tensor effective
frequency
\(\omega_{T,n}^2=q_n^2-a''/a\), with dotted reference levels indicating
\(q_n^2=n^2-1\). Panel (b) shows the implemented scalar canonical frequency
\(\omega_{S,n}^2\), again with dotted \(q_n^2\) reference levels. Panel (c)
shows the ratio \(\omega_{S,n}^2/q_n^2\). For sufficiently large \(n\), this
ratio remains close to unity through the bounce region, showing that the
closed-harmonic eigenvalue term dominates there. The lowest closed harmonics
are therefore the modes most sensitive to the bounce. The plotted vertical
range is clipped for readability in the far tails; no perturbation evolution
data are changed by this diagnostic. The direct low-\(n\) scan is a
conservative infrared regularity test rather than a direct computation of the
observable CMB spectrum.
}
\label{fig:mode_frequency_diagnostic}
\end{figure}

Fig.~\ref{fig:mode_frequency_diagnostic} illustrates why the direct
evolution focuses on the infrared sector. The lowest closed harmonics are the
ones for which the time-dependent effective potentials associated with the
bounce compete most strongly with the closed-harmonic eigenvalue term. For
sufficiently large \(n\), the ratio \(\omega_{S,n}^2/q_n^2\) remains close to
unity through the bounce region, so these modes remain adiabatic there; their
subsequent turning-point behavior is instead associated with the ordinary
inflationary freeze-out regime. The low-\(n\) scan is therefore a conservative
regularity test of the bounce-sensitive sector rather than an attempt to
compute the full observable CMB spectrum directly.

\subsection{Curvature perturbation and reported spectrum}
\label{app:zeta_and_power}

The curvature perturbation reported by the code is reconstructed from the evolved
mode via
\begin{equation}
 \zeta_n = \left(\frac{H}{\dot\phi}\right)_{\rm reg} Q_n,
\label{eq:zeta_appendix}
\end{equation}
using the same threshold-and-interpolation regularization strategy described
above. This definition is appropriate for the inflation-era freeze check, which
is the physical scalar diagnostic used in the body of the paper.

The default reported scalar spectrum is
\begin{equation}
 \mathcal P_s(n)
 =
 \frac{n(n^2-1)}{2\pi^2}\,|\zeta_n|^2
\label{eq:Ps_appendix}
\end{equation}
evaluated at the chosen output time. In the present work, the direct
closed-universe mode evolution is used as an infrared regularity diagnostic of
the bounce-plus-inflation background rather than as a brute-force derivation of
the observable astronomical-$n$ pivot spectrum.

\subsection{Numerical robustness checks}
\label{app:robustness}

A focused robustness sweep was performed over representative modes
$n=3,5,10,30,60$ by varying the interpolation threshold, the adiabatic-window
criterion, and the initialization point within the first valid WKB window.
These checks were designed to test whether the scalar-sector conclusions depend
sensitively on implementation choices rather than on the background itself.
The results are summarized in Table~\ref{tab:robustness}.

\begin{table*}[t]
\caption{Summary of scalar-sector robustness checks for representative modes
$n=3,5,10,30,60$. Across all tested variations, no mode failures or
$\Ucal_n$ pathologies were found, regular bounce passage was preserved,
and the qualitative freeze pattern was unchanged.}
\label{tab:robustness}
\begin{ruledtabular}
\begin{tabular}{p{0.22\textwidth}p{0.20\textwidth}p{0.14\textwidth}p{0.14\textwidth}p{0.22\textwidth}}
Test variation & Range tested & Failures / ambiguities & $\Ucal_n$ pathology & Main effect \\ \hline
Regularization threshold $\delta$
& $10^{-12}$--$10^{-16}$
& None
& None
& Negligible impact on reported results. \\

Adiabaticity criterion $\epsilon_{\rm ad}$ and minimum admissible window length
& $\epsilon_{\rm ad}=10^{-2}$--$10^{-1}$; window length $=3,5,7$
& None
& None
& Final representative-mode powers vary at the percent level in the worst cases; qualitative conclusions unchanged. \\

Initialization point in first valid WKB window
& earliest / midpoint / latest
& None
& None
& Percent-level shifts in final low-$n$ powers; freeze pattern unchanged. \\
\end{tabular}
\end{ruledtabular}
\end{table*}

Quantitatively, the robustness sweep confirms the same freeze pattern found in
the baseline run: modes $n=10,30,60$ are frozen at the few-$\times 10^{-3}$
level over the last five e-folds before $\epsilon_1=1$, while the deepest
tested infrared modes retain a stable residual drift, with $n=5$ at the
two-percent level and $n=3$ at the several-percent level. This residual behavior is therefore a feature of the
deepest tested infrared sector rather than an artifact of a particular
threshold or initialization choice.

\subsection{Homogeneous shear estimate}
As a compact first-pass anisotropy check, one may model a small homogeneous
shear contribution by an effective energy density
\(
\rho_\sigma\propto a^{-6}
\).
Normalizing this at the bounce by
\(
\varepsilon_\sigma\equiv \rho_{\sigma,b}/\rho_b
\),
with
\(
\rho_b=3/a_b^2
\)
for the explicit background, one has
\[
\rho_\sigma(t)=\varepsilon_\sigma \rho_b\left(\frac{a_b}{a(t)}\right)^6.
\]
Since along the exact closed-FRW branch
\(
\rho_{\rm tot}(t)=3(H^2+a^{-2})\ge 3a^{-2}
\),
it follows that
\[
\frac{\rho_\sigma}{\rho_{\rm tot}}
\le
\varepsilon_\sigma\left(\frac{a_b}{a}\right)^4
\le
\varepsilon_\sigma,
\qquad
\frac{\rho_\sigma}{1/a^2}
=
3\varepsilon_\sigma\left(\frac{a_b}{a}\right)^4
\le
3\varepsilon_\sigma.
\]
Thus the effective homogeneous shear is maximized at the bounce and is diluted
on both sides of it. In particular, if
\(
\varepsilon_\sigma\ll 1
\),
the explicit branch remains shear-subdominant through the bounce and the shear
becomes utterly negligible on the slow-roll inflationary branch. This is only
a compact homogeneous estimate not be be confused with a full nonlinear
anisotropic analysis.

\section{Geodesic completeness of the explicit background}
\label{app:direct_completeness}

The perturbation analysis uses a finite numerical background window extending
from the contracting branch through the end of inflation. Geodesic
completeness, however, is a statement about the full homogeneous solution, so
the relevant check must be performed on the same explicit background continued
beyond that finite window.

For the explicit homogeneous model used in this work, the compact-support
reheating switch removes the residual plateau tail exactly on the far right, so
the full implemented potential reaches the stabilized minimum at
\begin{equation}
\phi_{\min}=6.0,
\qquad
V(\phi_{\min})=\Lambda\simeq 4.75\times 10^{-121},
\label{eq:Vmin_appendix}
\end{equation}
so the corresponding de Sitter scale is
\begin{equation}
H_\Lambda=\sqrt{V(\phi_{\min})/3}\simeq 3.98\times 10^{-61}.
\label{eq:Hl_appendix}
\end{equation}

On the expanding side, the post-inflationary scalar exhibits a long oscillatory
matter-like transient, but the exact future asymptotics cannot remain
matter-like forever. Once the oscillation energy above the minimum redshifts
away, the future branch approaches de Sitter with
\begin{equation}
H(t)\to +H_\Lambda,
\qquad
a(t)\sim A_+ e^{H_\Lambda t}
\qquad (t\to+\infty),
\label{eq:future_dS_appendix}
\end{equation}
for some positive constant \(A_+\).

On the contracting side, a constraint-preserving reversed-time continuation of
the same explicit homogeneous solution shows that
\begin{equation}
H(t)\to -H_\Lambda,
\qquad
\phi(t)\to \phi_{\min},
\qquad
\dot\phi(t)\to 0
\qquad (t\to-\infty),
\label{eq:past_fields_appendix}
\end{equation}
so the past branch is asymptotically contracting de Sitter,
\begin{equation}
a(t)\sim A_- e^{-H_\Lambda t}
\qquad (t\to-\infty),
\label{eq:past_dS_appendix}
\end{equation}
for some positive constant \(A_-\).

The null and timelike completeness criteria for FRW backgrounds are controlled
by the divergence of
\begin{equation}
\int a(t)\,dt,
\qquad
\int \frac{a(t)}{\sqrt{a(t)^2+1}}\,dt
\label{eq:completeness_integrals_appendix}
\end{equation}
on the past and future ends~\cite{Lesnefsky:2022fen,Easson:2024fzn}. Since \(a(t)\) grows exponentially on both ends,
the null integral diverges immediately. Likewise, because \(a(t)\to\infty\) as
\(t\to\pm\infty\), one has
\begin{equation}
\frac{a(t)}{\sqrt{a(t)^2+1}}\to 1,
\end{equation}
so the timelike integral also diverges on both ends.

Thus the explicit curvature-bounce background studied in this paper is null-
and timelike-complete in the sense of these FRW completeness integrals.
\newpage

\bibliography{simple_early_universe_refs_final}

\end{document}